\documentclass[fleqn,usenatbib,useAMS]{mnras}

\usepackage[dvipsnames]{xcolor}
\usepackage{tikz,hyperref}
\usepackage{ulem}
\definecolor{lime}{HTML}{A6CE39}
\DeclareRobustCommand{\orcidicon}{
	\begin{tikzpicture}
	\draw[lime, fill=lime] (0,0) 
	circle [radius=0.13] 
	node[white] {{\fontfamily{qag}\selectfont \tiny ID}};
	\draw[white, fill=white] (-0.0625,0.095) 
	circle [radius=0.007];
	\end{tikzpicture}
	\hspace{-2mm}
}

\foreach \x in {A, ..., Z}{\expandafter\xdef\csname orcid\x\endcsname{\noexpand\href{https://orcid.org/\csname orcidauthor\x\endcsname}
			{\noexpand\orcidicon}}
}

\usepackage{newtxtext,newtxmath}

\usepackage{booktabs}

\usepackage[T1]{fontenc}
\usepackage{ae,aecompl}

\usepackage{graphicx}	
\usepackage{amsmath}	
\usepackage{booktabs}
\usepackage{times}
\input{epsf}

\title[H$_2$ in the Owl Nebula]
{H$_2$ molecular gas in the old planetary nebula NGC\,3587}

\author[G. Ramos-Larios et al.]{
G.\,Ramos-Larios\thanks{E-mail:\,gerardo@astro.iam.udg.mx}$^{1\orcidA}$, M.~A.\,Guerrero$^{2\orcidB}$, J.~A.~Toal\'{a}$^{3\orcidC}$, S.\,Akras$^{4\orcidD}$ and X.\,Fang$^{5\orcidE}$ 
\\
$^{1}$Instituto de Astronom\'{i}a y Meteorolog\'{i}a, CUCEI, Universidad de Guadalajara, Av. Vallarta 2602, Col. Arcos Vallarta, 44130 Guadalajara, Mexico\\
 $^{2}$Instituto de Astrof\'{\i}sica de Andaluc\'{\i}a, IAA-CSIC, C/Glorieta de la Astronom\'{\i}a s/n, 18008 Granada, Spain\\
$^{3}$Instituto de Radioastronom\'{i}a y Astrof\'{i}sica, UNAM, Ant. Carretera a P\'{a}tzcuaro 8701, Ex-Hda. San Jos\'{e} de la Huerta, Morelia 58089, Mich., Mexico\\
$^{4}$Institute for Astronomy, Astrophysics, Space Applications and Remote Sensing, National Observatory of Athens, GR 15236 Penteli, Greece\\
$^{5}$National Astronomical Observatories, Chinese Academy of Sciences, 20A Datun Road, Chaoyang District, Beijing, China
}

\date{\today}

\pubyear{2023}

\begin{document}
\label{firstpage}
\pagerange{\pageref{firstpage}--\pageref{lastpage}}
\maketitle

\begin{abstract}
The acquisition of high-quality deep images of planetary nebulae (PNe) has allowed the detection of a wealth of small-scale 
features, which highlight the complexity of the formation history and physical processes shaping PNe. 
Here we present the discovery of three groups of clumps embedded within the nebular shell of the evolved PN NGC\,3587, the Owl Nebula, that had escaped previous detections.  
The analysis of multi-wavelength GEMINI GMOS, NOT ALFOSC, Aristarchos Andor optical, CFHT WIRCam and {\it Spitzer} IRAC and MIPS infrared (IR) images indicates that these clumps are formed by material denser and colder than the surrounding nebula, with a notable content of molecular H$_2$, but negligible or null amounts of dust.  
The presence of H$_2$-rich pockets embedded within the ionized shell of this evolved PN is suggestive of the survival of high-density condensations of material created at the onset of the PN stage. 
 
\end{abstract}




\section{Introduction}\label{introduction}
\label{sec:intro}

Molecules are commonly found in the interstellar medium (ISM), frequently associated with star forming regions, dense and vast molecular clouds, and the circumstellar environment around evolved stars, to name a few. 
Among the large number of molecules in the ISM, one of the most abundant \textbf{is} molecular hydrogen (H$_2$). 
Its formation occurs through reactions on the surface of dust grains, which act as catalyst to absorb the energy excess that is released when a new H$_2$ molecule is formed, surviving to photo-dissociation in not too dense environments. 
Multiple transitions of this molecule can be detected at  infrared (IR) wavelengths. 
The near-IR H$_2$ 1$-$0 S(1) $\lambda$2.122 $\mu$m emission line is the brightest and most easily detected, although the H$_2$ 2$-$1 S(1), H$_2$ 1$-$0 S(2), H$_2$ 1$-$0 S(0), H$_2$ 1$-$0 S(3), and H$_2$ 2$-$1 S(3) near-IR transitions can be also detected.  
In the mid-IR, H$_2$ 0$-$0 transitions from the S(2) to the S(7) lines emission have also been observed, with the H$_2$ 0$-$0 S(3) $\lambda$9.66 $\mu$m line being among the brightest.

These emission lines are routinely detected in planetary nebulae (PNe).  
Since the pioneer work of \citet[]{1984A&A...130..151I}, many succeeding near-IR imaging and spectroscopic H$_2$ surveys of PNe have been published \citep[][and references therein]{1988MNRAS.235..533W,1995ApJS..100..159L,1996ApJ...462..777K,1999ApJS..124..195H,2000ApJS..127..125G,2017MNRAS.470.3707R,2018MNRAS.479.3759G}, as well as mid-IR ones \citep[e.g.,][]{2016MNRAS.459..841M,2017ApJS..231...22O,2022NatAs...6.1421D}. 
The detector technological advances and increasing telescope apertures have played an important role in the detection of new H$_2$-emitting PNe. 
Indeed, many PNe that were previously undetected in H$_2$ surveys \citep[e.g., A\,66, NGC\,6543, NGC\,7009, NGC\,7662 or M\,2-48,][and references therein]{1988MNRAS.235..533W,1996ApJ...462..777K} are now known to present H$_2$ emission associated with low-ionization structures \citep[e.g., K\,4-47, NGC\,6543, NGC\,7009, and NGC\,7662,][]{2017MNRAS.465.1289A,2018ApJ...859...92F,2020MNRAS.493.3800A}, the so-called LIS \citep[][]{2001ApJ...547..302G}, or with dense knots and clumps that are embedded within ionized material in ring-like structures \citep[e.g., NGC\,650-51,][]{2013MNRAS.429..973M}.

Among the PNe that exhibit H$_{2}$ emission there seem to be two types: those that exhibit thousands of clumps \citep[group A; e.g., NGC\,2346 and NGC\,3132;][]{Manchado2015,2022NatAs...6.1421D} while others are detected to harbour a few H$_2$-emitting knots \citep[group B; e.g., NGC~7662, NGC~6543, K~4-47, NGC~6778, NGC~3242, NGC~6826, NGC~6818, NGC~7354 and Hu~1-2;][]{2018ApJ...859...92F}. Group A is characterized by old PNe with ages $>$7000 yr and H$_{2}$ clumps with sizes of $[1-4]\times10^{15}$ cm \citep{Akras2020Galaxies}, whereas PNe in group B are younger ($<$2000 yr) and have clump sizes larger than $10^{16}$ cm.

Here we report the discovery of a number of clumps embedded within ionized material detected in optical low-excitation ionic and near-IR H$_2$ molecular line emission in NGC\,3587, a.k.a.\ the Owl Nebula, which were undetected in early studies \citep{1996ApJ...462..777K}. 
NGC\,3587 is a triple-shell PN consisting of an elliptical inner shell with the characteristic "owl eyes and beak", a round envelope, and an outer halo \citep[]{2003AJ....125.3213G}. 
The latter is asymmetrical, with a bow-shock like shape along the direction of motion of the central star revealing the interaction of the nebula with the ISM.
Several works outline precisely its physical structure observed through many emission lines \citep{1985MNRAS.217..539S}, including a complete analysis of the density distribution \citep{2000AJ....120.2661C} and a morpho-kinematic modeling that describes accurately the dynamics of this highly evolved PN and highlight the presence of a series of internal cavities \citep{2018MNRAS.479.3909G}. Interestingly, NGC\,3587 is an evolved PNe, with an estimated kinematic age $\simeq$10,400 yr and a linear size for its main nebula $\simeq$0.86 pc \citep{2003AJ....125.3213G} at the {\it Gaia} distance of 810$\pm$30 pc. That is, it should be part of group A of the H$_{2}$-emitting PNe, but as we will demonstrate its number of H$_{2}$-emitting clumps is not large.

\begin{figure}
\includegraphics[width=\columnwidth]{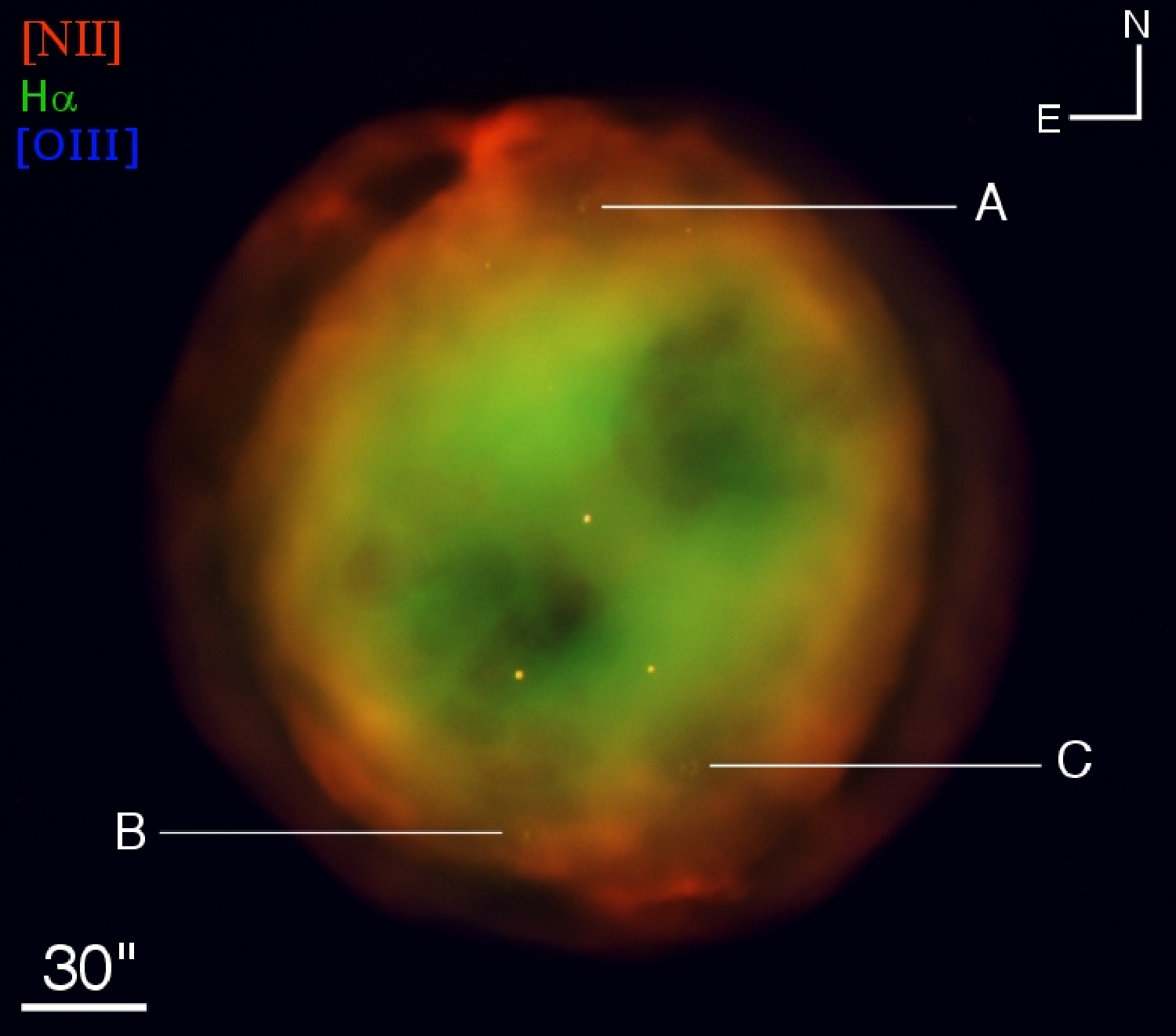}
\caption{
NOT ALFOSC colour-composite picture of NGC\,3587 of [N~{\sc ii}] (red), H$\alpha$ (green) and [O~{\sc iii}] (blue) images (see details in Sec.~\ref{sec:NOT}). 
The picture reveals three barely visible small discrete regions, labelled as A, B and C, that contrast with the large-scale filaments and bright and dark patches of the main nebula.
}
\label{fig:1}
\end{figure}

In order to investigate the nature of the newly discovered molecular hydrogen clumps and its spatial correlation with the ionized nebula and possible dust content, we have collected a large amount of high-quality observational material in the optical and IR range.
A general and brief description of the available observations is presented in Section~\ref{observations}, with the results described in Section~\ref{results}. 
The discussion is presented in Section~\ref{discussion} and the final conclusions in Section~\ref{conclusions}.

\section{Observations}\label{observations}

\subsection{Optical Imaging}

In the following we present the details of the optical images. A summary of the optical observations is presented in Table~\ref{tab:imaging}.

\subsubsection{Nordic Optical Telescope}
\label{sec:NOT}

The ALhambra Faint Object Spectrograph and Camera (ALFOSC) at the 2.56-m Nordic Optical Telescope (NOT) in the Observatorio de Roque de los Muchachos (ORM, La Palma, Spain) was used to obtain a series of images of NGC\,3587 in the [N~{\sc ii}], H$\alpha$ and [O~{\sc iii}] emission lines on 2018 June 20.  
The detector used was a 2K$\times$2K e2v CCD with plate scale 0.21 arcsec~pix$^{-1}$ and a field of view (FoV) of 6$\farcm$5$\times$6$\farcm$5.  
The central wavelengths and bandwidths of the narrow-band filters are 
$\lambda_\mathrm{c}$=6583\,\AA\ and $\Delta\lambda$=13\,\AA\ for [N~{\sc ii}], 
$\lambda_\mathrm{c}$=6565\,\AA\ and $\Delta\lambda$=13\,\AA\ for H$\alpha$, and 
$\lambda_\mathrm{c}$=5007\,\AA\ and $\Delta\lambda$=30\,\AA\ for [O~{\sc iii}].    
The spatial resolution achieved during the observations, as 
determined from the FWHM of stars in the FoV, was $\simeq$0.85~arcsec. 
Three individual frames with integration times of 600\,s were taken for the [N~{\sc ii}] and H$\alpha$ filters, leading to total exposure times of 1800\,s, whilst three frames with integration times of 200\,s were used for the [O~{\sc iii}] for a total exposure time of 600\,s.

The NOT images were bias-subtracted and flat-fielded using twilight flats employing standard {\sc iraf}\footnote{The Image Reduction and Analysis Facility, {\sc iraf}, is distributed by the National Optical Astronomy Observatory, which is operated by the Association of Universities for Research in Astronomy (AURA) under cooperative agreement with the National Science Foundation.} routines \citep{Tody1993}. 
A NOT ALFOSC colour-composite picture of the Owl Nebula in the light of  H$\alpha$, [N~{\sc ii}] and [O~{\sc iii}] is presented in Figure~\ref{fig:1}.

\begin{table}\centering
\setlength{\columnwidth}{0.1\columnwidth}
\setlength{\tabcolsep}{1.0\tabcolsep}
\caption{Details of the optical and IR imaging.}
\begin{tabular}{lllcc}
\hline
\multicolumn{1}{l}{Telescope} &
\multicolumn{1}{l}{Instrument} &
\multicolumn{1}{l}{Filter} & 
\multicolumn{1}{c}{$\lambda_\mathrm{c}$} & 
\multicolumn{1}{c}{$\Delta\lambda$} \\ 
\multicolumn{1}{c}{} &
\multicolumn{1}{c}{} & 
\multicolumn{1}{l}{} & 
\multicolumn{1}{c}{(\AA)} & 
\multicolumn{1}{c}{(\AA)} \\ 
\hline
NOT     & ALFOSC  & [O~{\sc iii}]     & 5007 & 30 \\
        &         & H$\alpha$         & 6567 & 13 \\
        &         & [N~{\sc ii}]      & 6588 & 13 \\
GEMINI	& GMOS    & $g$ (G0301)       & 4744 & 1700 \\
        &         & H$\alpha$ (G0310) & 6576 & 140 \\
        &         & $r$ (G0303)       & 6298 & 1700 \\
Aristarchos & Andor ikon-L  & [O~{\sc i}]     & 6304 & 30 \\
        &         & [S~{\sc ii}]      & 6727 & 40 \\
CFHT    & WIRCam  & $K_\mathrm{c}$             & 22180 & 330 \\
        &         & H$_2$             & 21220 & 320 \\
        &         & Br$_\gamma$       & 21660 & 300 \\
{\it Spitzer} & IRAC    & 3.6               & 35500 & 7500 \\
        &         & 4.5               & 44930 & 19015 \\
        &         & 8.0               & 78720 & 29050 \\
        & MIPS    & 24                & 240000 & 50000 \\
\hline
\end{tabular}
\label{tab:imaging}
\end{table} 

\begin{figure*}
\includegraphics[width=0.95\linewidth]{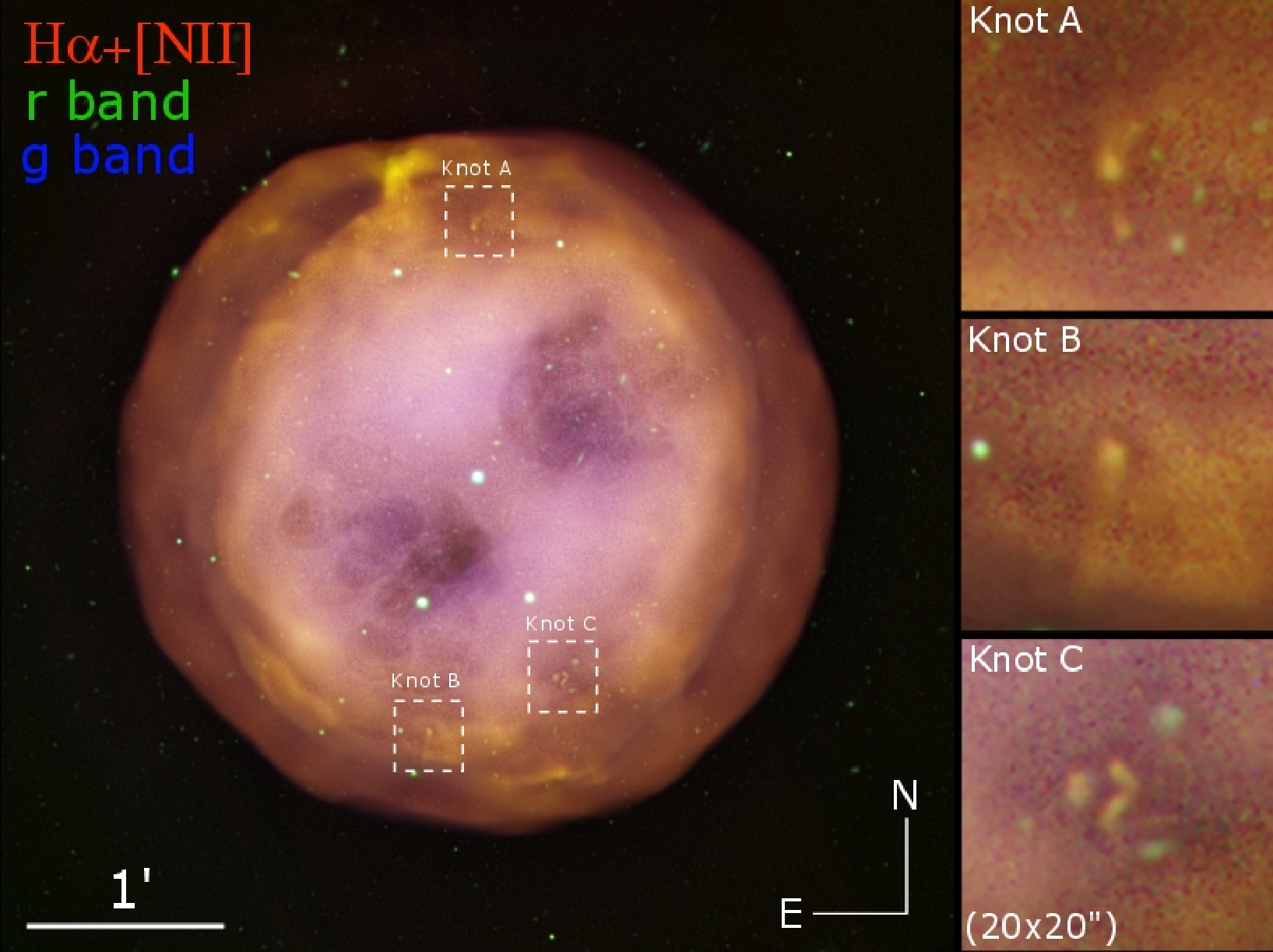}
\caption{
(left) Gemini North GMOS high-resolution colour-composite picture of NGC\,3587 in the light of the H$\alpha$+[N~{\sc ii}] (red), $r$-SDSS (green), and $g$-SDSS (blue) filters, and (right) insets of the 20\arcsec$\times$20\arcsec squared boxes overlaid on the picture around the position of the areas of lumpy material denoted as A, B and C in Figure~\ref{fig:1}. 
Their ``unusual'' shapes are emphasized in these zoomed views, with a bow-shock-like structure apparently associated with feature B.
The image has been processed using unsharp masking techniques to emphasize small details in the main nebula. 
}
\label{fig:2}
\end{figure*}

\subsubsection{Gemini North Telescope}

Very high-angular-resolution optical images of NGC\,3587 were acquired at the Gemini North Observatory with the Gemini North Multi-Object Spectrographs (GMOS-N) on April 15, 2009 in the narrow-band H$\alpha$+[N~{\sc ii}] G0310 filter and broad-band $r$-SDSS (G0303) and $g$-SDSS (G0301) filters. 
The exposure times were 8$\times$90\,s, 16$\times$30\,s and 16$\times$30\,s, respectively, on 2$\times$2 binning.
The unbinned detector array consisted of three 2048$\times$4608 EEV chips for a total of 6144$\times$4608 pixels arranged in a row with $\sim$0.5\,mm gaps with pixel size of 13.5$\times$13.5 $\mu$m, plate scale of 0\farcs0727~pixel$^{-1}$, and FoV of 5\farcm5$\times$5\farcm5. The seeing during the observations, as 
determined from the FWHM of stars in the FoV, was $\simeq$0.65~arcsec.
The Gemini images were processed  employing {\it gmosaic} and {\it gemtools} tasks and standard {\sc iraf} routines. 
A Gemini North GMOS colour-composite picture of the Owl Nebula is presented in Figure~\ref{fig:2}.

\subsubsection{Aristarchos Telescope}

The 2.3m Aristarchos telescope at the Helmos Observatory (Greece) was used to obtain a series of images in the [O~{\sc i}] $\lambda$6300 and [S~{\sc ii}] $\lambda\lambda$6716,6731 emission lines on 2023 July 17.  
The camera used was an Andor ikon-L with a CCD e2v CCD42-40 detector, which included 2048$\times$2048 pixels$^2$ with 13.5~$\mu$m pixel size, resulting to a spatial scale of 0\farcs16~pix$^{-1}$ and a FoV of 5$\farcm$5$\times$5$\farcm$5. In this case, the resolution achieved during the observations, as 
determined from the FWHM of stars in the FoV, was $\simeq$1.75~arcsec.
The central wavelengths and bandwidths of the  filters are $\lambda_\mathrm{c}$=6304\,\AA\ and $\Delta\lambda$=30\,\AA\ for [O~{\sc i}] and
$\lambda_\mathrm{c}$=6727\,\AA\ and $\Delta\lambda$=40\,\AA\ for [S~{\sc ii}]. 
The [O~{\sc i}]~6300\AA~and [S~{\sc ii}]~6716+6731\AA\ images were obtained with exposure times of 2$\times$900~sec and 900~sec, respectively. 

Standard {\sc iraf} routines were followed for the images processing, which includes bias subtraction, flat fielding correction, bad-pixel screening, and cosmic-ray removal. 
Figure~\ref{fig.stavros} displays a 65$\times$65 arcsec$^2$ section of the [O~{\sc i}] image towards the northern region of the Owl Nebula.

\subsection{IR Imaging}

\subsubsection{CFHT WIRCam}

The Wide-field InfraRed Camera (WIRCam) \citep[]{2004SPIE.5492..978P} is an state-of-art near-IR wide-field imaging instrument mounted at the prime focus of the 3.6m Canada-France-Hawaii Telescope (CFHT). 
WIRCam consists of a mosaic of four HAWAII2-RG detectors, each containing 2048$\times$2048 pixels, with a sampling of 
0.3 arcsec~pix$^{-1}$ and an impressive FoV of 20$\farcm$0$\times$20$\farcm$0.  
A series of exposures in the narrow-band filters of $K_\mathrm{c}$ ($\lambda_\mathrm{c}$=2.218\,$\mu$m and $\Delta\lambda$=0.033\,$\mu$m), H$_2$ ($\lambda_\mathrm{c}$=2.122\,$\mu$m and $\Delta\lambda$=0.032\,$\mu$m), and Br$\gamma$ ($\lambda_\mathrm{c}$=2.166\,$\mu$m and $\Delta\lambda$=0.030\,$\mu$m) were acquired on 2013 December 22 and 25 (program ID 13BS01, P.I.\ Yong Zhang). 
Sixteen images with individual exposure times of 65\,s for H$_2$, whilst twelve and eight frames with exposures of 60\,s for the Br$\gamma$ and $K_\mathrm{c}$ filters respectively were combined in order to improve the signal-to-noise (S/N) ratio. 

\begin{figure}
\centering
\includegraphics[width=0.75\linewidth]{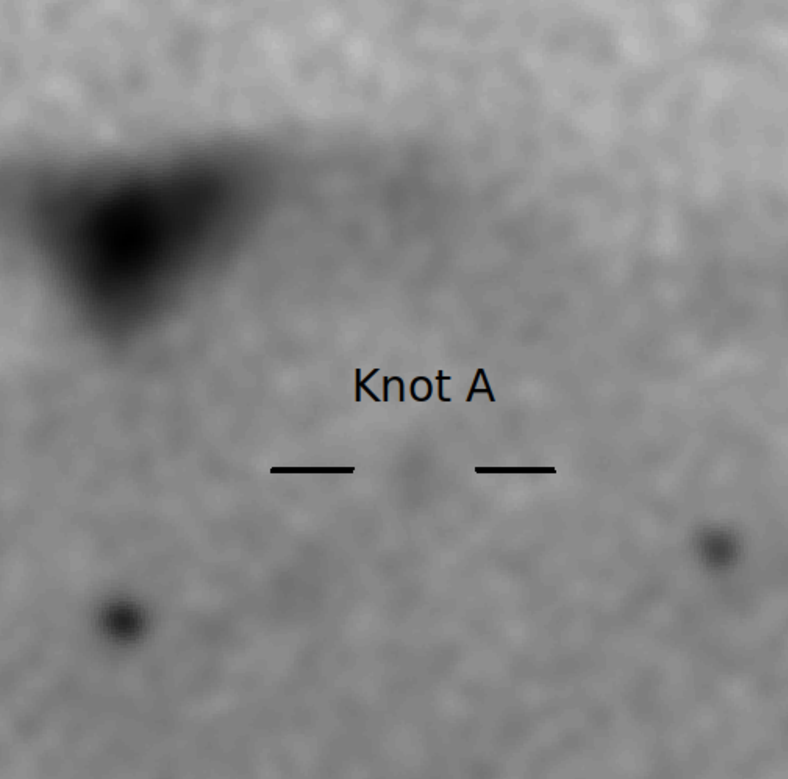}
\hspace*{0.01cm}
\includegraphics[width=0.75\linewidth]{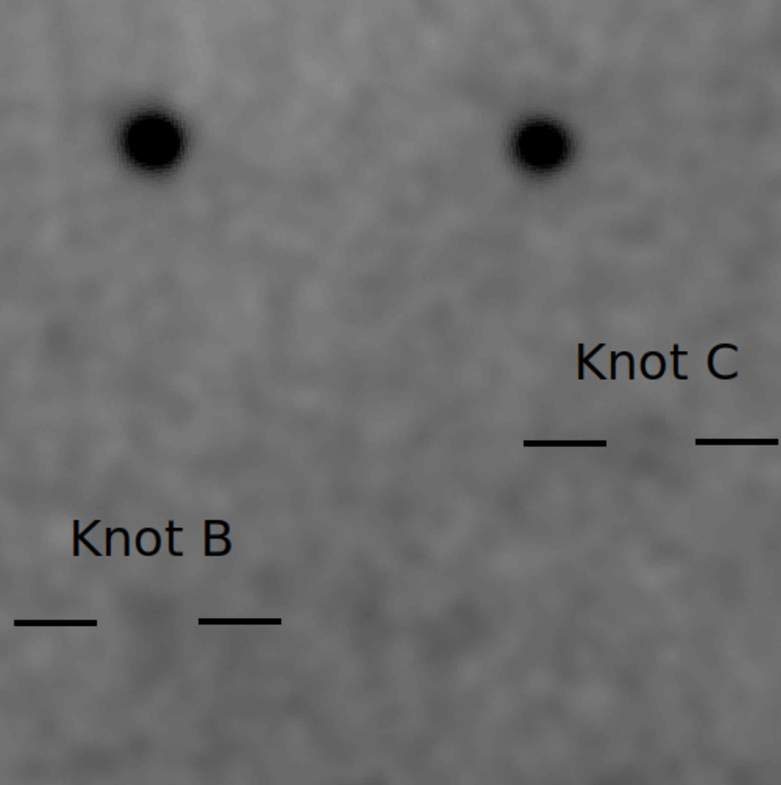}
\caption{Aristarchos narrow-band [O~{\sc i}] $\lambda$6300 grayscale images of NGC\,3587 centred at clumps A (top), and B and C (bottom). 
The horizontal lines indicate the position of the clumps.
The size of the images is 65$\times$65 arcsec$^2$. 
North is up, east to the left.}
\label{fig.stavros}
\end{figure}

\begin{figure}
\begin{center}
	\includegraphics[width=0.805\columnwidth]{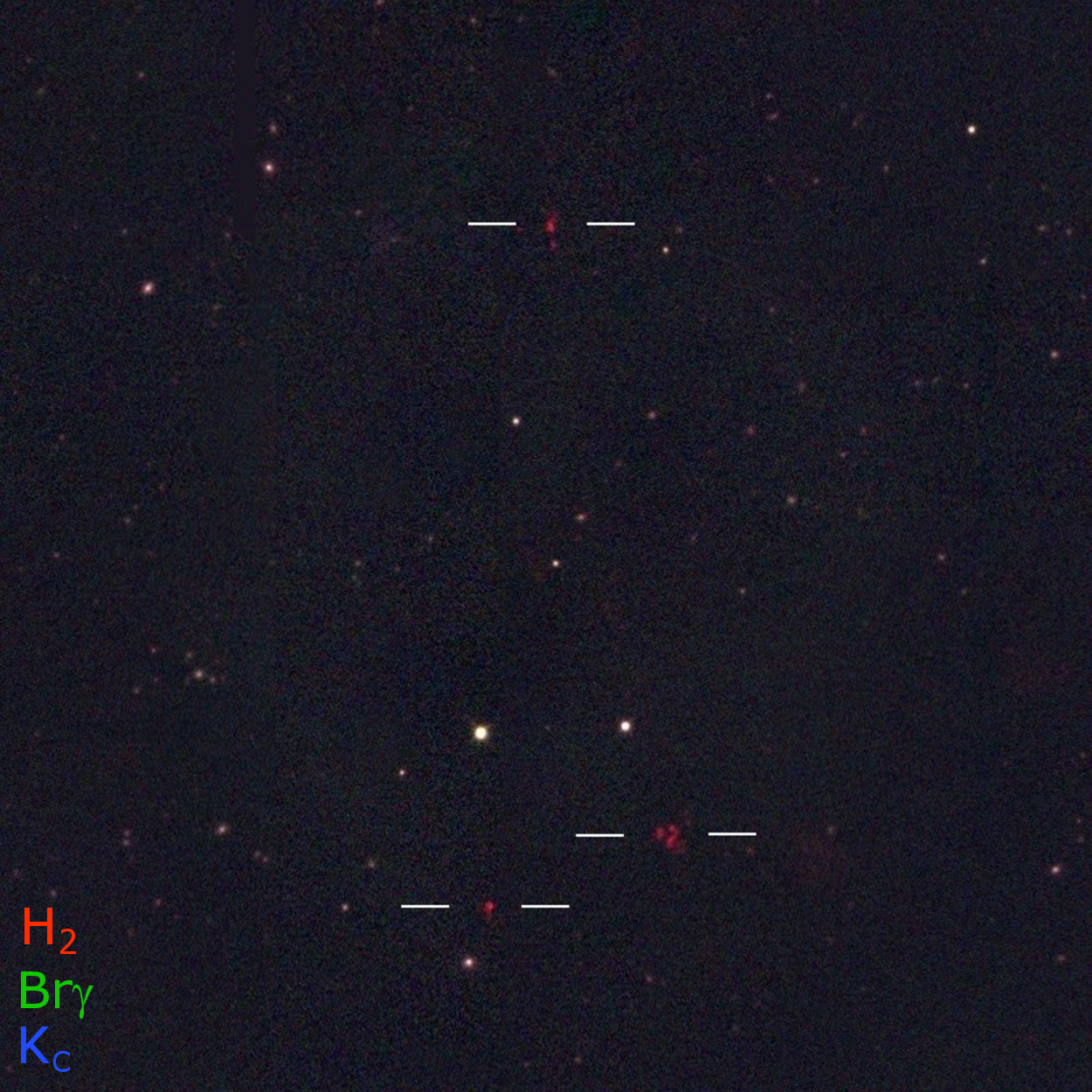}\\
	\includegraphics[width=0.805\columnwidth]{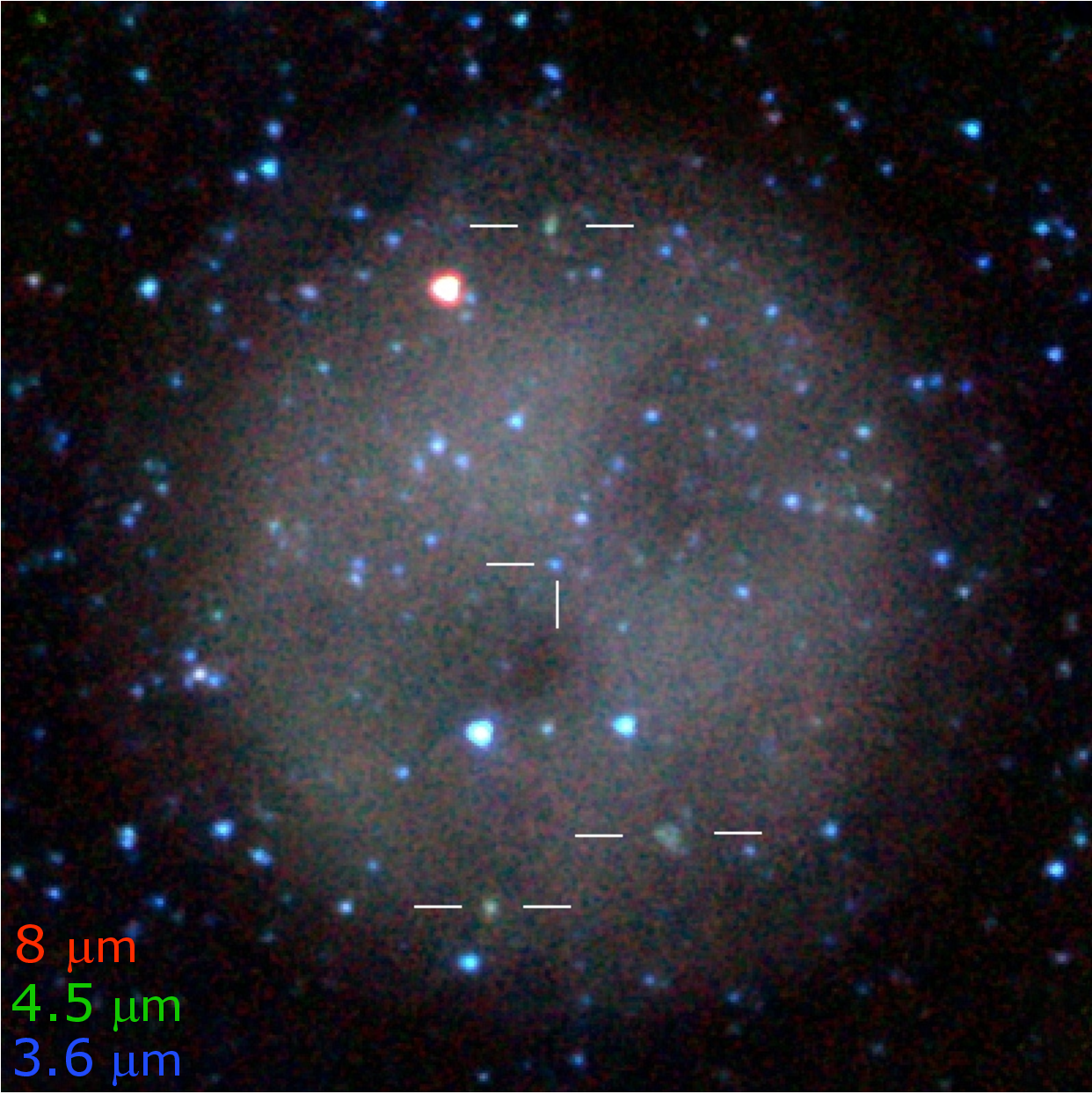}\\
	\includegraphics[width=0.805\columnwidth]{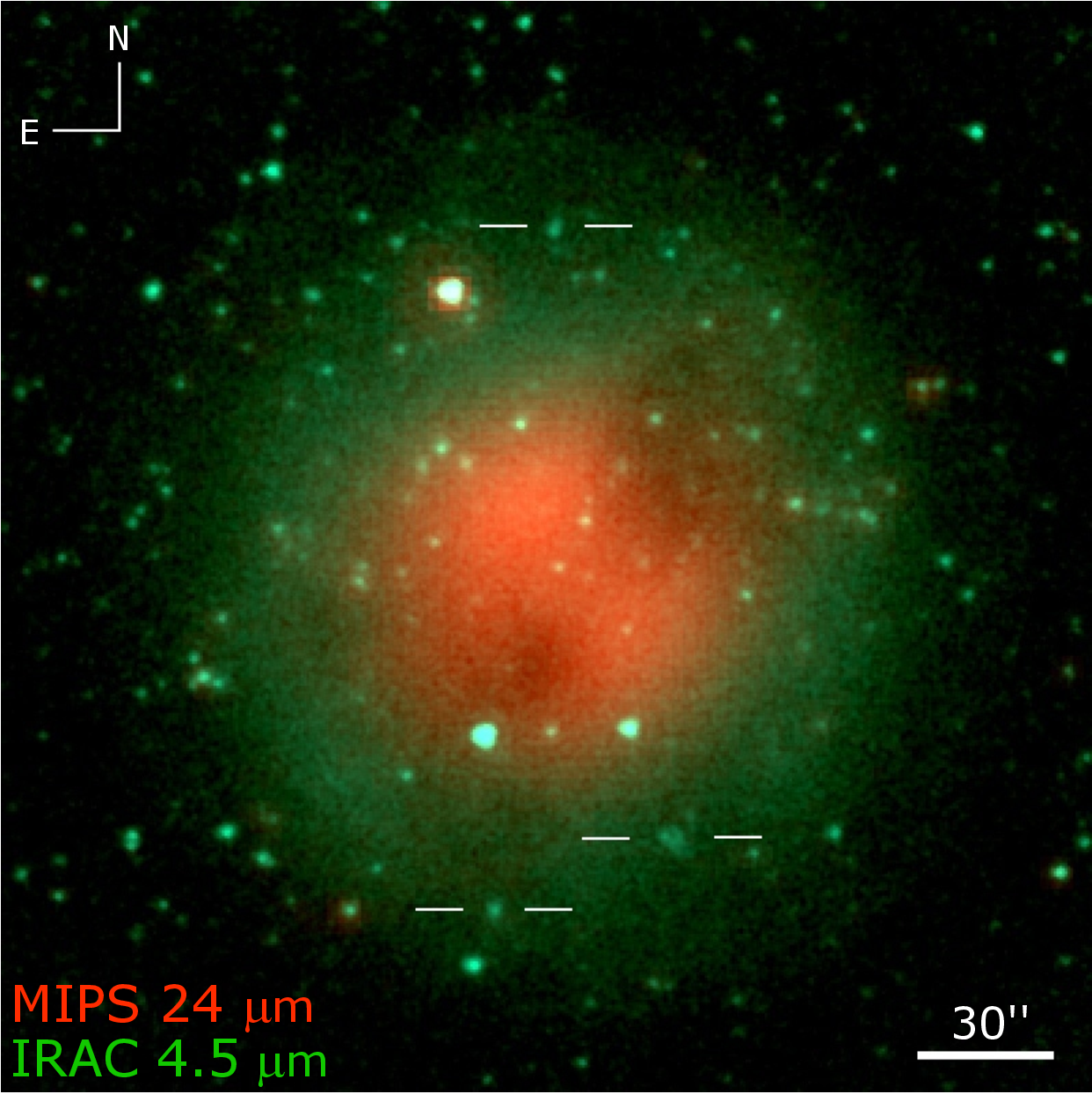}
\end{center}
\caption{
Near- and mid-IR colour-composite pictures of NGC\,3587 obtained combining CFHT H$_2$ (red), Br$\gamma$ (green), and $K_\mathrm{c}$ (blue) images (top), 
{\it Spitzer} IRAC 3.6 $\mu$m (blue), 4.5 $\mu$m (green), and 8 $\mu$m (red) images (middle), and 
{\it Spitzer} MIPS 24 $\mu$m (red) and IRAC 4.5 $\mu$m (green) images (bottom).  
All pictures share the same scale and FoV.  
The position of the progenitor star is marked at the centre of the middle panel, whilst horizontal lines in the three panels show the position of H$_2$ molecular hydrogen as clearly denoted in the top panel. 
Note the conspicuous lack of nebular Br$\gamma$ emission in the CFHT colour-composite picture (top panel).
}
\label{fig:3}
\end{figure}

Similarly to the optical imaging obtained at the NOT, all the WIRCam images were reduced following standard IR procedure using {\sc iraf} routines, including dark subtraction, flat-field correction, basic cross-talk removal, and sky subtraction for the final combination of each set of images.
A colour-composite image of NGC\,3587 using the WIRCam narrow-band images is presented in Figure~\ref{fig:3}-top. \\

\begin{figure*}
\includegraphics[width=\columnwidth]{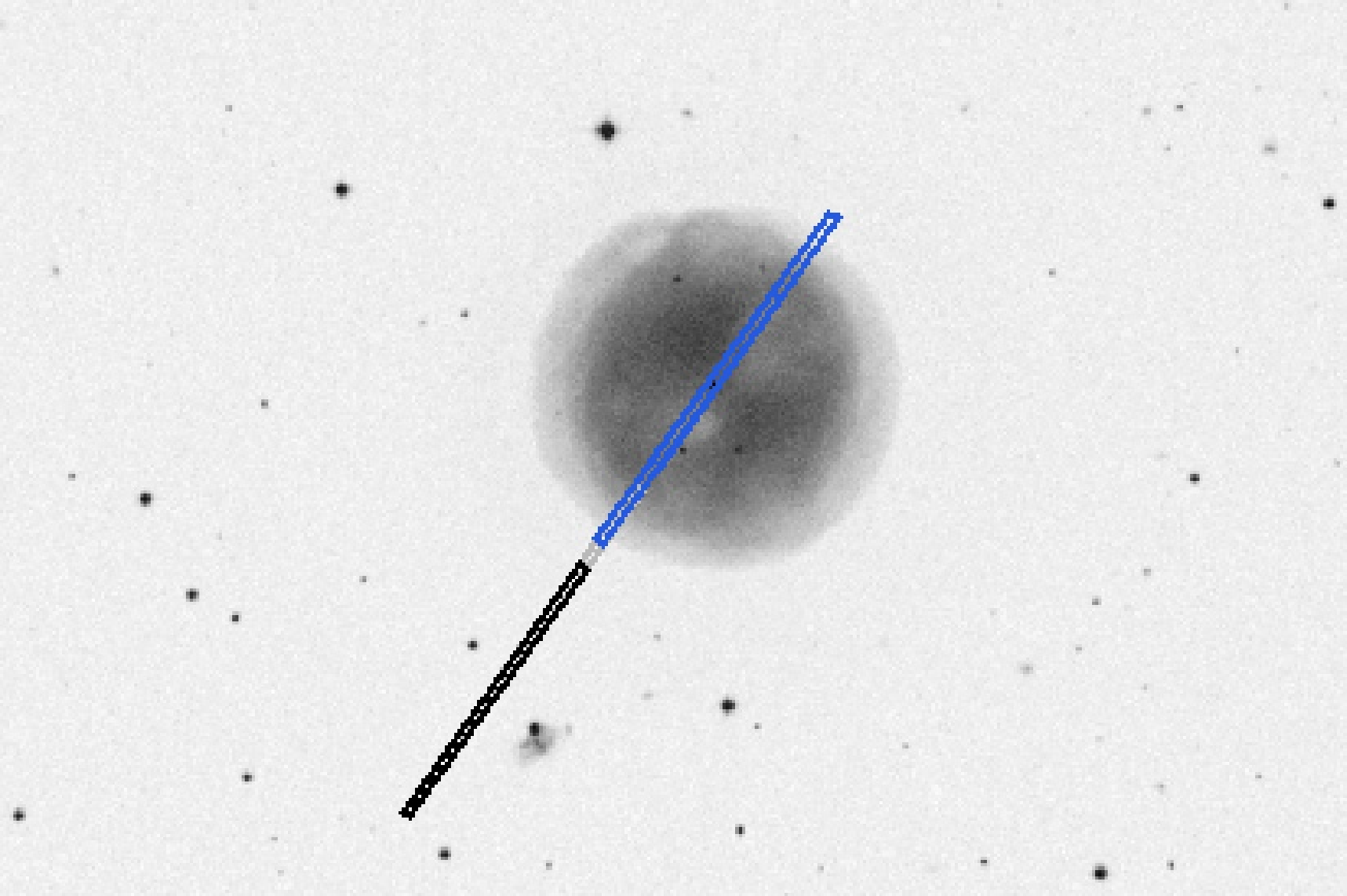}
\includegraphics[width=\columnwidth]{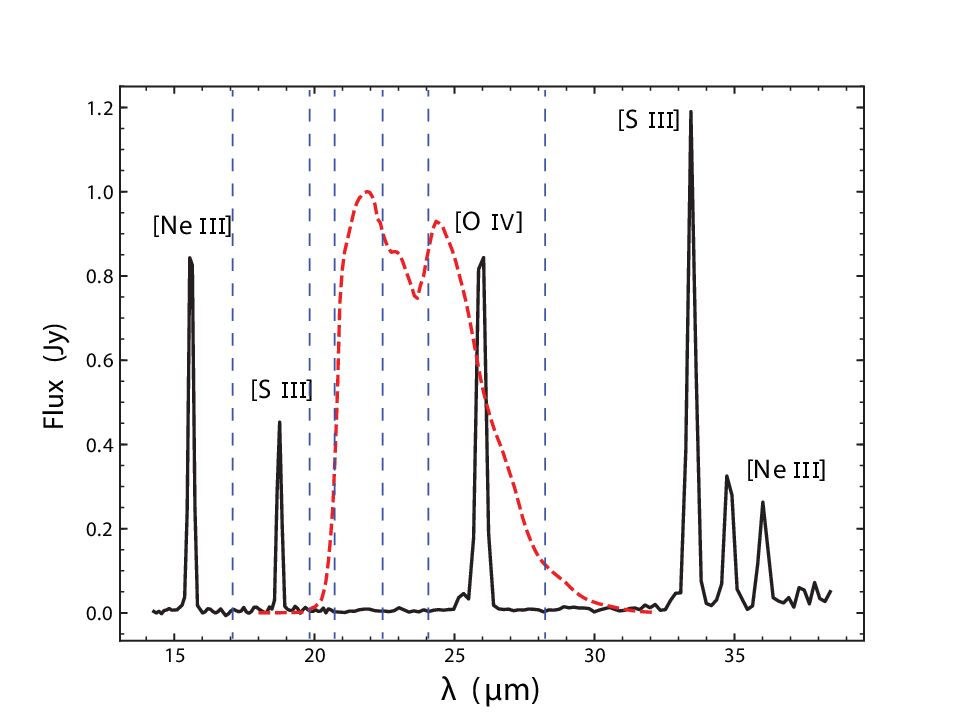}
\caption{
(left) Slit position for the {\it Spitzer} IRS LL modules (blue) and slit position for background subtraction (black). (right) {\it Spitzer} background subtracted spectra. 
The spectral region around 21 $\mu$m shows the overlap of the LL1+LL2 spectra. 
Main ionic lines are identified as well as some H$_2$ lines in the {\it S} and {\it O} transitions marked in dashed blue, like the H$_2$ 0$-$0 S(0) at 28.2188 $\mu$m and the H$_2$ 6$-$5 O(16) at 19.8082 $\mu$m among others.
The MIPS 24 $\mu$m transmission curve is also marked with a red dashed line.  
As mentioned in the text, unfortunately none of the available slits positions intersect the H$_2$ zones observed in the main nebula. 
}
\label{fig:spitzer_irs}
\end{figure*}

\subsubsection{Spitzer IRAC \& MIPS}

Images from the Infrared Array Camera \citep[IRAC;][]{2004ApJS..154...10F} on board {\it Spitzer} were obtained from the National Aeronautics and Space Administration/Infrared Processing and Analysis Center (NASA/IPAC) Infrared Science Archive (IRSA). IRAC is a four-channel camera capable for simultaneous imaging in broadband filters with central wavelengths 3.6, 4.5, 5.8 and 8.0 $\mu$m. 
The detector arrays of 256$\times$256 pixels in size use InSb for the two shorter-wavelength 3.6 and 4.5 $\mu$m channels, and Si:As IBC detectors for the two longer-wavelength 5.8 and 8.0 $\mu$m channels.

The set of images were acquired on 2004 April 20 as part of Program ID 68
(Studying Stellar Ejecta on the Large Scale using SIRTF-IRAC,
PI: Giovanni Fazio). The observations presented here make use only of the filters at 3.6 $\mu$m, 4.5 $\mu$m and 8 $\mu$m (see middle panel in Fig.~\ref{fig:3}). The more noisy 5.8 $\mu$m image was discarded for the final composition. \\

In addition, we also retrieved a 24~$\mu$m image of NGC\,3587 obtained with the Multiband Imaging Photometer \citep[MIPS;][]{2004ApJS..154...25R} on board {\it Spitzer}. 
These data were obtained on 2008 January 8 as part of Program ID 40953 (MIPS Survey of Dust Disks around Hot White Dwarfs, PI: You-Hua Chu). 
As the IRAC long-wavelength bands, the MIPS 24 $\mu$m camera uses the Si:As IBC detector array of 128$\times$128 pixels.
The MIPS 24~$\mu$m band is mainly dominated by dust continuum,
but also the [Ne~{\sc v}] at 24.3 $\mu$m and the [O~{\sc iv}] at 25.9 $\mu$m lines are strong candidates to MIPS fluxes in nebular components \citep[see][]{Chu2009}.

Colour-composite pictures of NGC\,3587 using the {\it Spitzer} IRAC 3.6, 4.5, and 8 $\mu$m images, and the {\it Spitzer} IRAC 4.5 $\mu$m and MIPS 24 $\mu$m images are presented in the middle and bottom panels of Figure~\ref{fig:3}, respectively.

\subsection{IR Spectra}

\subsubsection{{\it Spitzer} IRS}

Spectroscopic observations obtained with the {\it Spitzer} Infrared Spectrograph \citep[IRS;][]{2004ApJS..154...18H} were downloaded from the NASA/IPAC Infrared Science Archive. 
The instrument consists mainly of four separate modules, the Short-Low (SL) and Short-High (SH) ones, which provide low resolution ($R\sim$60--128) spectroscopic capabilities in the spectral range from 5.2 to 38 $\mu$m, and the Long-Low (LL) and Long-High (LH) modules, with moderate or high resolution ($R\approx$600) in the spectral range from 9.9 to 37.2 $\mu$m. 
The set of spectra used here were acquired on 2004 April 17 as part of Program ID: CAL IRS 1406/1406 (Program reserved for IRS calibration observations, PI: Calibration, IRS).

In this work we only present the IRS spectra obtained with the Long-Low LL1 and LL2 modules with suitable slits positions for background subtraction. 
The plate scale was 5.1~arcsec~pix$^{-1}$ and the aperture sizes for the LL1 and LL2 modules were 10.7$\times$168 arcsec$^{2}$ and 10.5$\times$168 arcsec$^{2}$, respectively.  
The slits had a position angle (PA) of $\approx$145$^\circ$. 
The spectra cover the spectral ranges 14--21.3 $\mu$m for LL2 and 19.5--38 $\mu$m for LL1 with a resolution R$\sim$57--126. 
All the IRS spectra were processed with the software package Cube Builder for IRS Spectral Mapping \citep[{\sc cubism};][]{2007PASP..119.1133S}, which is a tool for constructing spectral cubes, maps and arbitrary aperture 1D spectral extractions from sets of mapping mode spectra. 
The slit position and LL spectra are shown in Figure~\ref{fig:spitzer_irs}.  
None of the available slits cover the small H$_2$ reservoirs in the main nebula. 

As for the available SL observations, not presented here, a carefully examination of the spectra obtained with the SL1 and SL2 modules did not found any H$_2$ emission lines. 
Since these data lack of suitable positions for background subtraction, they were discarded for subsequent analysis.

\section{Results}\label{results}

The narrow- and broad-band optical and IR images of the Owl Nebula presented in the previous section contain information on the nebular emission on a wide excitation range. 
The overall ionization and physical structure of the Owl Nebula have been the subject of many detailed studies (see Sect.~\ref{sec:intro}).  
Here we will rather focus on three rather inconspicuous clumps first uncovered in the NOT ALFOSC H$\alpha$ and [N~{\sc ii}] images (see Fig.~\ref{fig:1}), subsequently confirmed in the Gemini GMOS H$\alpha$+[N~{\sc ii}] image (Fig.~\ref{fig:2}) and Aristarchos Andor [O~{\sc i}] images (Fig.~\ref{fig.stavros}).

These clumps are labeled as A, B, and C in the NOT ALFOSC H$\alpha$ and [N~{\sc ii}] and Gemini GMOS images presented in Figures~\ref{fig:1} and \ref{fig:2}. 
Clumps A and B are detected at an angular distance from the CSPN $\approx75^{\prime\prime}$ along PAs $\approx0^\circ$ and $\approx170^\circ$, respectively, i.e., they are almost at the same distance but along opposite directions from the CSPN.  
Meanwhile, clump C, along PA $\approx190^\circ$, is closer to the CSPN, at $\approx65^{\prime\prime}$. 
The clumps are thus projected onto the main nebular shell, which can be encircled within an ellipse $182^{\prime\prime}\times168^{\prime\prime}$ in size with its major axis oriented along PA $\approx135^\circ$  \citep{2003AJ....125.3213G}. 
Adopting a distance of 810~pc will locate these knots at a projected distance on the plane of the sky from the CSPN of 0.29 pc for knots A and B, and 0.26~pc for knot C. 
If truly contained within the main nebular shell, then clumps A and B would be within 25$^\circ$ of the plane of the sky, whereas clump C can reach up to 40$^\circ$.

The clumps are quite noticeable in the CFHT WIRCam H$_2$ and {\it Spitzer} IRAC images in the mid-IR 3.6 and 4.5 $\mu$m bands (see Fig.~\ref{fig:3}), 
much weaker in {\it Spitzer} IRAC images at 5.8 and 8.0 $\mu$m, and [O~{\sc i}], with clump A detected while clumps B and C are only marginally detected, and undetected in [O~{\sc iii}], [S~{\sc ii}], Br$\gamma$, $K_\mathrm{c}$, and in the mid-IR {\it Spitzer} 24 $\mu$m band. 
The lack of [O~{\sc iii}] emission and detection in [O~{\sc i}] seem to imply material of lower ionization, yet the [N~{\sc ii}] to H$\alpha$ ratio at their cores derived from the NOT ALFOSC images, $\approx$0.15, is actually lower than that of the surrounding nebular material, $\approx$0.4. 
The emission of the clumps is thus actually enhanced in the recombination line of H$\alpha$, but reduced in the collisionally excited lines of [O~{\sc iii}], [N~{\sc ii}], and [S~{\sc ii}].  
These line ratios  variations  can be attributed to higher electronic density and lower temperature. 
Furthermore, the detection of [O~{\sc i}] emission in clump A and marginally in clumps B and C clearly implies the presence of neutral gas and its strong link with molecular hydrogen likewise other microstructures (e.g. NGC\,7009, K\,4-47, among others).

Thus the detection of the clumps in near-IR H$_2$ narrow-band images, but unnoticeable emission in Br$\gamma$, allow to reasonably attribute the emission in the {\it Spitzer} IRAC bands to H$_2$ $\nu$ $=$ 0--0 transitions. Therefore, the clumps consist mostly of denser and colder ionized material with a significant molecular content.

We further inspected the {\it Spizer} IRS spectra to search for possible H$_2$ emission in the mid-IR range. The left panel of Figure~\ref{fig:spitzer_irs} shows the slit position of the IRS data with the source and background regions. A background-subtracted spectrum was extracted in a region covering the extension of the Owl Nebula and is shown in the right panel of Figure~\ref{fig:spitzer_irs}. 
We detect a collection of emission lines in the {\it Spitzer} IRS spectrum, which include [Ne\,{\sc iii}], [S\,{\sc iii}] and [O\,{\sc iv}], but none corresponding to any of the H$_2$ $\nu$=0--0 series.
The detection of emission lines of high excitation species is expected given the high surface temperature $T_{\rm eff}$ = 93.9$\pm$5.6 kK of its CSPN \citep{1999A&A...350..101N}.

\begin{figure}
\includegraphics[width=\columnwidth]{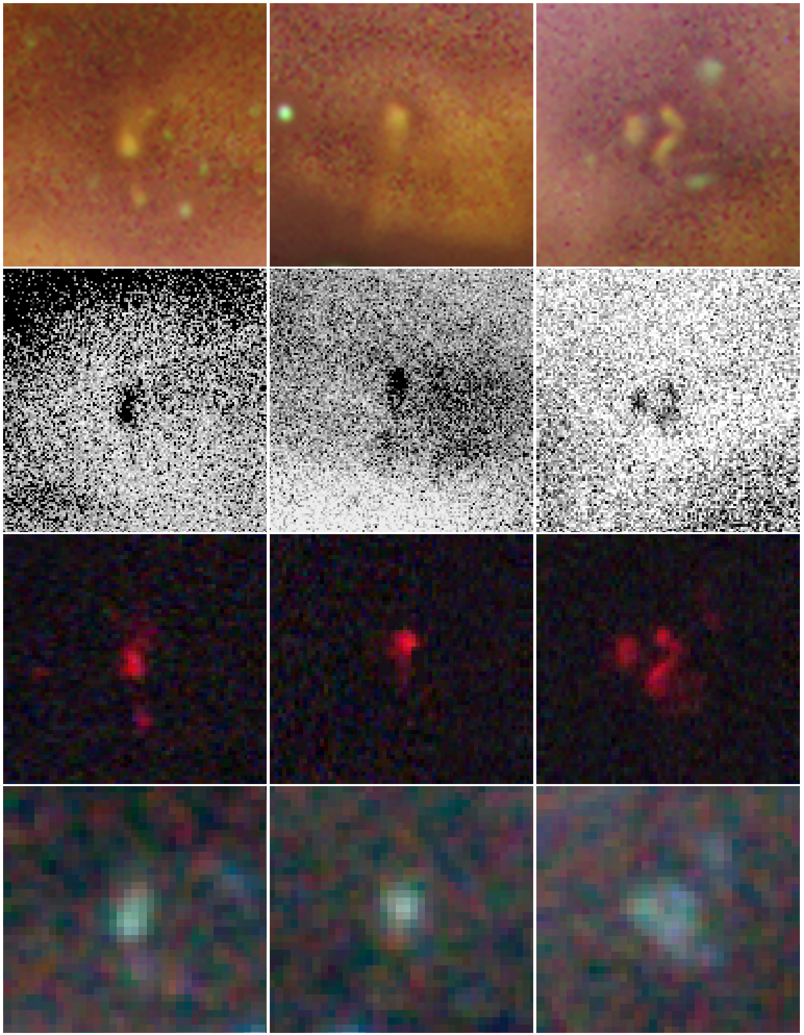}
\caption{
Colour-composite zoom-in views of the ionic low-excitation and molecular emission clumps A (left column), B (middle column), and C (right column) of NGC\,3587.  
From top to bottom, 
Gemini GMOS-N H$\alpha$+[N~{\sc ii}] (red), $r$-SDSS (green), and $g$-SDSS (blue) colour-composite picture, 
NOT ALFOSC [O~{\sc iii}]/H$\alpha$ ratio maps (where black corresponds to low values of this ratio), 
CFHT H$_2$ (red), Br$\gamma$ (green) and $K_\mathrm{c}$ (blue) colour-composite picture, and 
{\it Spitzer} IRAC 3.6 $\mu$m (blue), 4.5 $\mu$m (green), and 8 $\mu$m (red) colour-composite picture. 
All frames have the same spatial scale, with a field of view of 20\arcsec$\times$20\arcsec.
}
\label{fig:zoom}
\end{figure}

Zoom-in images of the clumps in the relevant emission line and broad-band images, as well as in the [O\,{\sc iii}]/H$\alpha$ ratio map, typically used to trace shocks in ionized structures and in particular in PNe \citep{Guerrero2013}, are presented in Figure~\ref{fig:zoom}.  
These images and ratio map reveal that clumps A, B, and C have different morphologies, although each is remarkably similar through the different images and ratio map, besides the varying spatial resolutions of these images. 
Clump A (left column in Fig.~\ref{fig:zoom}) consists of two compact knots aligned along a direction close but not coincident with the radial direction towards the CSPN.  
The innermost knot is fainter than the outermost one, which shows a diffuse and fainter broad tail-like structure extending outwards.   
Clump B (middle column in Fig.~\ref{fig:zoom}) has a compact head and a $\approx$25~arcsec long tail pointing outwards from the CSPN, with a possible broad bow-shock structure pointing inwards the nebula.  
Finally clump C (right column in Fig.~\ref{fig:zoom}) has a knotty, irregular appearance.  
Interestingly, the Gemini GMOS colour-composite picture of this clump reveals differences in colour, with the westernmost structures being brighter in the broad-band images, which are not apparent in the CFHT near-IR and {\it Spitzer} mid-IR colour-composite pictures. 
These westernmost structures may instead be associated with background galaxies, of which many are detected in the FoV with Gemini and {\it Spitzer} colours similar to these ones.

\section{Discussion}\label{discussion}

As mentioned in Section~\ref{sec:intro}, NGC~3587 is an evolved PN and its 
CSPN has a high UV flux, while the nebular density is generally low, with no high density structure\footnote{The structure of the Owl Nebula has been claimed to be bipolar, formed by a cylinder-like structure and a pair of cavities cleared out by one or even several fast outflows \citep[][]{2018MNRAS.479.3909G}, yet there is no evidence for a density contrast between these structures.} capable of providing  shelter for the survival of molecular material in this harsh environment. 
Evolved PNe with H$_2$ clumps usually have a large number of them (group A presented in Sect.~1), which is not the case for NGC\,3587. 
As for the physical size of the clumps, which is $\sim$1.2$\times$10$^{16}$~cm for an averaged angular radius of 1$^{\prime\prime}$, it is also larger than those of knots in evolved PNe, in the range of a few 10$^{15}$~cm.

The CFHT WIRCam and {\it Spitzer} IRAC IR imaging of NGC\,3587 (Fig.~\ref{fig:3}) confirms the presence of molecular material in three different locations.
Whilst no Br$\gamma$ emission is detected from ionized material in the CFHT WIRCam image, the presence of molecular hydrogen is quite obvious in Fig.~\ref{fig:3}-top and \ref{fig:zoom}. 
Moreover, the same features are nicely detected in bluish-green colour from the nebular background in the {\it Spitzer} IRAC images (Fig.~\ref{fig:3}-middle and \ref{fig:zoom}), suggesting that this latter features correspond to an excesses in the 3.6 and 4.5 $\mu$m bands.  
The emission in these bands can be attributed to the H$_2$ $\nu$ $=$ 0$-$0 forest transitions, namely S(9) to S(18), thus supporting the molecular content of these clumps.

Similar pockets of molecular material embedded within ionized material are being detected in an increasing number of PNe as the quality of their H$_2$ observations improves \citep[e.g.,][]{2013MNRAS.429..973M,2020MNRAS.493.3800A}. 
Maybe the most prominent example of these structures is the myriad of H$_2$-emitting knots embedded within NGC\,7293, the Helix Nebula \citep{2007AJ....133.2343O,2007MNRAS.382.1447M}.  
These show morphologies that can easily match that of the knots detected in NGC\,3587, as the cometary knots with tails (clump B) or multipeak knots (clump A) described by \citet{2009ApJ...700.1067M}.  
As the Owl Nebula, the Helix Nebula is an evolved PN with an estimated age of 10,000 yr \citep{1999ApJ...522..387Y,2008MNRAS.384..497M}.  
Its hot CSPN, with a high effective temperature of $T_{\rm eff}$ = 103.6$\pm$5.5 kK \citep[][]{1999A&A...350..101N}, confirms its evolved status.
The spatial distribution of the knots in the Helix Nebula, along an annular or ring-like structure, is otherwise notably different to that observed in the Owl Nebula. The inner H$_2$-rich clumps in the Helix Nebula seem to have been sculpted during the post-AGB evolution of its progenitor star. 
Thus clumps and filaments would have formed as a result of hydrodynamical instabilities created during the interaction between the (fast) post-AGB and the previously ejected (slow) AGB wind.

The molecular material found in NGC\,2346 provides another fair comparison with the clumps reported here in NGC\,3587. 
High spatial resolution H$_2$ images reveal the true physical structure nature of its molecular material, composed of knots and filaments \citep{Manchado2015}.   
The formation of these structures has been proposed by means of a mechanism of depressurization in the central bubble of NGC\,2346, which leads to the fragmentation of the shell into a network of knots and filaments. 
As the central star gets older and its UV flux declines, nebular material is not photoionized, surviving as molecular material.
Instead, the spatial distribution of the H$_2$ clumps in the Owl Nebula is more alike that of the H$_2$ knots located along the periphery of the envelope of NGC\,7662 \citep{2017MNRAS.465.1289A}.  

NGC\,7662 is much younger than the Owl Nebula, with a kinematic age $\approx$2,100 yr \citep{Guerrero_etal2004} at the revised {\it Gaia} distance of 1.7$\pm$0.9 kpc \citep{BailerJones2021}, which may explain the larger number of H$_2$ clumps than those observed in the Owl Nebula, where most clumps may have already dispersed.

Clump B in NGC\,3587 has the typical morphology of photoevaporated clumps seen in PNe and proplyds in the Orion Nebula \citep[see, e.g.,][]{GA2001}: a shocked-shell surrounding the dense clump produced by the shocked photoevaporation flow, the dense clump, and the tail. 
Other morphologies, as those exhibited by clumps A and C, can be shaped as well as the result of the complex dynamics produced by strong ionizing sources.  
Indeed, the radiative-magnetohydrodynamic 3D simulations presented by \citet{Henney2009} to follow the evolution of photoevaporative flows of an initially spherical clump result in a plethora of morphologies.

The origin of molecular material in old PNe is disputed, but it is accepted that it can form by the time the AGB envelope is ejected when the star's effective temperature is low, just before the PN is born. 
It is difficult, however, to reconcile the short time-scale of destruction of molecules on the environment of a PN \citep{Huggins2002} with the evolutionary stage of NGC\,3587.  
\citet{2014apn6.confE.117H} establishes two scenarios for the clumps survival: radiation and non-radiation. 
For the first case, they found that clumps photo-evaporate in shorter time-scales than those for the second case, where clumps are slowly destroyed by the wind. 
Moreover, 
\citet{2007ApJ...671L.137H} set down that the main destruction processes for the H$_2$ molecular material is due to photoionization by extreme ultraviolet radiation, which is relevant for the case of evolved PNe with molecular material.
Whatever the main mechanism might be, whether it involves winds or UV radiation, it is apparent that the clump structures would be survivors of molecular material.

Alternatively, it has been suggested that molecules and particularly H$_2$ can form at late times onto the surface of dust grains \citep{Aleman2004}.  
At any rate, the detection of clumps of material where H$_2$ molecules are found in evolved PNe implies that compact, high-density structures are able to survive the whole PN evolution (and even after the PN dispersal).

As for the Owl Nebula, the density of the clumps can be roughly estimated from the comparison of the H$\alpha$ flux of the clumps with the H$\alpha$ flux of the nebula, as density scales with the root square of the flux of the H~{\sc i} Balmer lines.  
Since the nebular density is $\simeq$600 cm$^{-3}$ \citep{2000AJ....120.2661C} and the H$\alpha$ flux of the clumps is $\approx$1.5 times that of the nebula, as determined from an area of similar angular size of the clumps and the nebular H$\alpha$ average surface brightness, the density of the clumps is $\approx$730 cm$^{-3}$. 
 
The higher electron density of the clumps in NGC\,3587 contrasts with the typically lower density of LIS with respect to their surrounding nebulae \citep[e.g. ][]{2016MNRAS.455.930A,2023MNRAS.518.3908M}. 
At any rate, the mass of clumps A, B, and C, derived from their density and volume, would be $\approx$10$^{-6}$ M$_\odot$, which is a tiny fraction of the ionized nebular mass.

\section{Conclusions}\label{conclusions}

The observations of NGC\,3587 here presented reveal for the first time isolated small-scale H$_2$-emitting features. 
The emission from these elusive structures would arise from clumps of material denser than the environment around them, thus allowing molecular material either to survive the whole PN evolution or to condense onto dust grains.  

Unfortunately, none of the MIR spectroscopic data available ranging between 5.2 and 38 $\mu$m registers the small H$_2$ zones observed, which would have made it possible to discern the true nature of the clumps by means of the H$_2$
rotational emission lines, deriving the Boltzmann distribution to calculate the H$_2$ excitation temperature.

This evolved and well studied object can now be considered a new member of PNe with H$_2$ content.  
The persistence of this molecular material in small pockets is intriguing.

\section*{Acknowledgements} 

The authors thank the anonymous referee for comments and suggestions that improved the presentation of the results. GRL acknowledges support from Consejo Nacional de Ciencia y Tecnolog\'{\i}a (CONACyT) grant 263373 and Programa para el Desarrollo Profesional Docente (PRODEP) (Mexico).  
MAG also acknowledges support of grant PGC2018-102184-B-100 co-funded with FEDER funds.
JAT thanks support from the project UNAM PAPIIT IA101622. Based on observations obtained with WIRCam, a joint project of CFHT, Taiwan, Korea, Canada, France, at the Canada-France-Hawaii Telescope (CFHT) which is operated by the National Research Council (NRC) of Canada, the Institut National des Sciences de l'Univers of the Centre National de la Recherche Scientifique of France, and the University of Hawaii. 
The Nordic Optical Telescope, operated on the island of La Palma jointly by Denmark, Finland, Iceland, Norway and Sweden, in the Spanish Observatorio del Roque de los Muchachos of the Instituto de Astrof\'{\i}sica de Canarias.
Based on observations made with the 2.3 m Aristarchos telescope, Helmos Observatory, Greece, which is operated by the Institute for Astronomy, Astrophysics, Space Applications and Remote Sensing of the National Observatory of Athens, Greece.
This research uses data obtained through the Telescope Access Program (TAP), which is funded by the National Astronomical Observatories, Chinese Academy of Sciences, and the Special Fund for Astronomy from the Ministry of Finance. 
This work is based in part on observations made with the Spitzer Space Telescope,  which  is  operated  by  the  Jet  Propulsion  Laboratory,  California  Institute  of  Technology  under  a  contract  with NASA. This work has made extensive use of NASA's Astrophysics Data System (ADS).

\section*{Data availability}

The data underlying this work will be shared on reasonable request to the corresponding author.











\bsp	
\label{lastpage}
\end{document}